# Exosphere Modeling of Proxima b: A Case Study of Photochemical Escape with a Venus-like Atmosphere

Yuni Lee[1,2,3,4], Chuanfei Dong[5,6], and Valeriy Tenishev[7]

**Abstract**

Exoplanets orbiting M-dwarfs within habitable zones are exposed to stellar environments more extreme than that terrestrial planets experience in our Solar System, which can significantly impact the atmospheres of the exoplanets and affect their habitability and sustainability. This study provides the first prediction of hot oxygen corona structure and the associated photochemical loss from a 1 bar $CO_2$-dominated atmosphere of a Venus-like rocky exoplanet, where dissociative recombination of $O_2^+$ ions is assumed to be the major source reaction for the escape of neutral O atoms and formation of the hot O corona (or exospheres) as on Mars and Venus. We employ a 3D Monte Carlo code to simulate the exosphere of Proxima Centauri b (PCb) based on the ionosphere simulated by a 3D magnetohydrodynamic model. Our simulation results show that variability of the stellar wind dynamic pressure over one orbital period of PCb does not affect the overall spatial structure of the hot O corona but contributes to the change in the global hot O escape rate that varies by an order of magnitude. The escape increases dramatically when the planet possesses its intrinsic magnetic fields as the ionosphere becomes more extended with the presence of a global magnetic field. The extended hot O corona may lead to a more extended H exosphere through collisions between thermal H and hot O, which exemplifies the importance of considering nonthermal populations in exospheres to interpret future observations.

[1] Corresponding author yuni.lee@nasa.gov

[2] University of Maryland Baltimore County, Baltimore, MD, USA
[3] Center for Research and Exploration in Space Science and Technology II, Baltimore, MD, USA
[4] NASA Goddard Space Flight Center, Planetary Environments Laboratory, code 699, Greenbelt, MD, USA
[5] Princeton Plasma Physics Laboratory, Princeton University, Princeton, NJ, USA
[6] Department of Astrophysical Sciences, Princeton University, Princeton, NJ, USA
[7] Department of Climate and Space Sciences and Engineering, University of Michigan, Ann Arbor, MI, USA

# 1. Introduction and Motivation

The processes by which the current status of planetary atmospheres evolved from their earlier conditions are complex. Loss of atmosphere is one of the ways how a planet balances the energy from its host star absorbed by its atmosphere [*Catling and Zahnle*, 2009; *Lammer et al.*, 2009]. Unless blasted away by impacts of asteroids or comets, atmospheric gases escape to space through several different processes categorized by the source mechanisms: thermal escape processes (*i.e.,* hydrodynamic and Jeans escape) [*e.g., Tian et al.*, 2015] and nonthermal escape processes (*e.g.,* photochemical escape, sputtering, and ion outflow) [*e.g., Laneuville et al.*, 2020]. For an escape to occur, an atmospheric particle must pass through the background atmospheric gases and overcome the planet's gravitational force. Hence, escape processes primarily depend on the composition of an atmosphere, the size and mass of a planet (escape velocity), and the interaction between the host star and the planet. Depending on these characteristics, the dominant escaping gases and the escape processes vary for different planets.

In our Solar System, several distinct escape processes for atmospheric constituents from different rocky planets have been explored by many missions, including Mars Atmosphere and Volatile and EvolutioN (MAVEN), Mars Express (MEX), Earth's Cluster, and Venus Express (VEX). These missions have improved our current understanding of terrestrial planetary atmospheres, plasma environments, and various escape processes for the present epoch, shedding light on the evolution of planetary atmospheres. What we have learned from the planets in the Solar System led us to ponder (1) how planetary exospheres form under different stellar environments in extrasolar systems, (2) how planetary (upper) atmospheres evolve under different stellar wind and interplanetary magnetic field conditions, (3) what roles of the magnetic fields of a planet play on the atmospheric loss, and (4) what escape processes play an important role in shaping the atmospheres.

The recent discoveries of rocky exoplanets orbiting their host stars allow us to broaden our perspective of atmospheric evolution. The distribution of those planets in the parameter space of planetary radius and orbital period implies that atmospheric loss plays a crucial role in shaping the atmosphere and determining the planet's fate [*Kite and Barnett*, 2020]. Most known exoplanets, including those with Earth-like size and mass in the Solar System's neighborhood, found to orbit M-dwarfs in their habitable zones, dominating the general occurrence rates of rocky planets in our galaxy [*Chabrier and Baraffe*, 2000; *Dressing and Charbonneau*, 2015; *Tuomi et al.,* 2019; *Winters et al.,* 2015]. As the atmospheres of the exoplanets orbiting M-dwarfs are exposed to more extreme environments than the Solar System bodies experience, the observations of such exoplanets raise questions about the atmospheric loss mechanisms and the sustainability of their atmospheres.

The closest super-Earth exoplanet (1.27 Earth masses), Proxima Centauri b (PCb), is a planet orbiting an M-dwarf host (M5.5), Proxima Centauri, which lies in the classical 'habitable zone' [*Anglada-Escudé et al.,* 2016]. Because of its proximity to our Solar System and similarity to the Earth and Venus in size, potential habitability and atmospheric evolution of PCb has been of considerable interest to the planetary community [*e.g., Bonfils et al.*, 2013; *Dong et al.,* 2017a; *Meadows et al.,* 2018; *Tuomi et al.,* 2014]. Due to the extreme stellar environment in the close-in habitable zone [*e.g., Dong et al.,* 2017a; *Garraffo et al.*, 2016], the atmosphere of PCb can undergo severe erosion through several different processes.

Many modeling efforts have contributed to improving our understanding of exoplanetary interaction with the host star and the resulting evolution of their atmospheres. In those studies, the loss process of atmospheric gases is often assumed to be hydrodynamic escape [*e.g., Erkaev et al.,* 2013; *Lopez et al.,* 2012, 2017; *Owen and Jackson*, 2012]. Some more advanced modeling techniques have been applied to understand the dynamics of planetary envelops and their interaction with host stars [*e.g., Cohen et al.,* 2015; *Dong et al.*, 2017b, 2018a, 2019, 2020; *Ekenback et al.,* 2010; *Erkaev et al.,* 2007; *Kislyakova et al.,* 2013, 2014; *Airapetian et al.,* 2020]. Nevertheless, there have been only a limited number of studies on secondary volatile atmospheres and relevant escape processes (*e.g.,* nonthermal processes) that are important for the habitability and sustainability of planetary atmospheres.

This study uses PCb as a benchmark planet for understanding atmospheric loss from Venus-like exoplanets with a 1 bar $CO_2$-dominated atmosphere. For the first time, we examined the loss of neutral oxygen atoms from PCb via photochemical escape (*i.e.,* a nonthermal escape process) and formation processes of resulting exoplanetary hot atomic coronae or exospheres. In the Solar System, the photochemical escape of neutral O atoms is the most dominant atmospheric loss mechanism on Mars at the present epoch. Despite the current cold and arid conditions, there is substantial observational and theoretical evidence that liquid water once existed on Mars [*e.g., Carr*, 1986, 2007; *Dong et al.,* 2018b; *Jakosky and Phillips*, 2001; *Jakosky et al.,* 2018]. Along with the H loss, many studies have shown that the loss of O to space directly reflects the water inventory in the Martian atmosphere [*e.g., Chassefiere et al.,* 2013; *Jakosky et al.,* 2018; *Lammer et al.,* 2003; *McElroy*, 1972]. This study aims to provide one possible scenario of atmospheric loss on PCb and any Venus-like exoplanets orbiting M-dwarfs. This study utilizes our modeling capability established and validated throughout the MAVEN mission [*e.g., Dong et al.,* 2015a, 2018c; *Lee et al.,* 2018, 2020] for the exosphere, ionosphere, and magnetosphere of Mars as a baseline to further extend our understanding in the parameter space of planetary and stellar environments for rocky exoplanets.

In Section 2, we provide the descriptions for our models and the simulation setup. Section 3 describes the formation process of hot O corona at PCb. The simulation results for the cases considered are shown in Section 4, and we discuss our results and conclusions in Section 5.

**2. Model Description and Simulation Setup**

The simulations for this study are conducted via our coupled modeling framework (see Figure 1, see Sections 2.1-2.3 for details). We employ our existing models for Venus and Mars [*e.g., Ma et al.*, 2013; *Dong et al.,* 2017a; *Lee et al.,* 2015a], assuming the atmospheric composition of PCb is similar to that of Venus. The coupling procedures of the models have been established for previous investigations of the atmospheric loss at Venus and Mars. For this study, we integrated our models for the exosphere, ionosphere, and magnetosphere to consider the full macroscopic interaction between the stellar wind and the upper atmosphere of PCb.

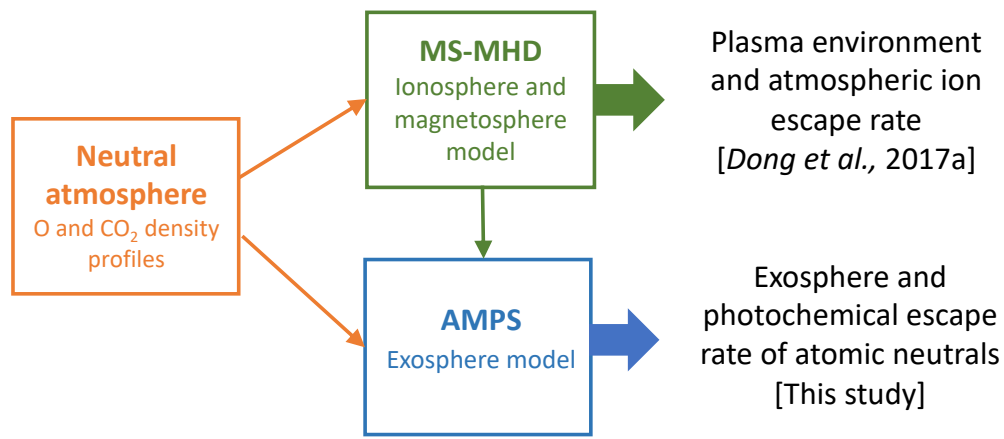

**Figure 1.** A diagram illustrating our coupled modeling framework. The thin arrows indicate the direction of the model data flow. For example, MS-MHD and AMPS take the density profiles of O and $CO_2$ in the thermosphere as their inputs. The thick arrows show the final products of the models.

**2.1. Adaptive Mesh Particle Simulator (AMPS): Exosphere Model**

This study's primary numerical tool is the state-of-the-art 3D Adaptive Mesh Particle Simulator (AMPS) [*Tenishev et al.*, 2008, 2011, 2013]. We use the AMPS code to investigate the formation and structure of the exosphere of PCb and the photochemical escape of neutral constituents. The AMPS code is developed within the frame of the Direct Simulation Monte Carlo (DSMC) method [*Bird*, 1994], which simulates the ensemble of model particles and captures the collisional dynamics of the gas species in a rarefied gas flow regime. As the code is generic, it has been used widely and well-tested for various kinetic problems in our Solar System, including the gas and dust distributions in cometary comas [*e.g., Tenishev et al.*, 2008, 2011; *Combi et al.*, 2012; *Bockelée-Morvan et al.*, 2015; *Fougere et al.*, 2016a, b], the plumes of Enceladus [*e.g., Waite et al.*, 2006; *Tenishev et al.*, 2010, 2014], and the exosphere of the Moon [*e.g., Tenishev et al.*, 2013]. For this study, we use our application to the Mars' exosphere [*e.g., Valeille et al.,* 2010b, 2010b; *Lee et al.,* 2014a, 2014b, 2015a, 2015b]. The code for the Mars application has been used to investigate the photochemical escape of neutral atomic species and to provide predictions of the exospheric structure for the current Mars mission. More detailed technical specifications of the AMPS code and its other capability are given in *Tenishev et al.* [2008, 2013, 2021].

AMPS utilizes a Cartesian adaptive mesh based on the Adaptive Mesh Refinement (AMR) technique [*Berger and Colella*, 1989]. The grid size is variable and optimized to provide appropriate descriptions of the regions in the atmosphere with essential features. The nominal upper boundary of the computational domain is set to about a few planetary radii (6 PCb radii for this study), which allows minimization of computational time and resource without underestimating or overestimating the escape rates of the model particles.

**2.2. Multi-species Magnetohydrodynamic Model (MS-MHD): Magnetosphere Model**

The 3D Block Adaptive Tree Solar-wind Roe Upwind Scheme (BATS-R-US) multi-species MHD (MS-MHD) model [*Tóth et al.,* 2012] provides planetary ionospheric profiles to AMPS. The MS-MHD model for PCb in this study [*Dong et al.,* 2017a] was adapted from its Venus application [*Ma et al.,* 2013]. The model solves the multi-species single-fluid MHD equations, where there are four continuity equations for $H^+$, $O^+$, $O_2^+$, and $CO_2^+$ with one momentum and energy equations. The ionosphere is self-consistently computed starting from the model's inner boundary at ~100 km altitude, which incorporates detailed ion dynamics and photochemical processes resulting from the interaction with the stellar wind and radiation. The model grid has a horizontal resolution of 3º × 3º and a varying radial resolution from 5 km to a few thousand kilometers, which is constructed to capture the critical physics in different altitude regions without overwhelming the entire computation [*Dong et al.,* 2017a].

**2.3. Coupling Procedure and Model Parameters**

It is important to note that the modeled exoplanet in this study does not necessarily represent PCb as the actual atmospheric conditions of PCb are not yet fully known. Our simulations for PCb are designed to provide one of many possible scenarios for the PCb's atmosphere and any other similar exoplanets orbiting M-dwarfs with planetary conditions that resemble our modeled exoplanet. More importantly, this study aims to build a basis of our understanding of exoplanets like PCb by applying knowledge learned from our Solar System.

The integration procedure of the two models is illustrated in Figure 1. The models are integrated such that one model provides the background conditions to the other. For this study, the MS-MHD model is simulated first to describe the ionosphere and magnetosphere of PCb for the cases considered. The AMPS model then adopts the global description of the ionosphere (*i.e.,* densities and temperature of the major ions and electrons) from the MS-MHD model as input to simulate the PCb's exosphere. For consistency, both MS-MHD and AMPS employ the same stellar and planetary parameters in this study. The model parameters for AMPS, the densities of thermosphere and ionosphere species (O, $CO_2$, $O_2^+$, and $e^-$), and temperatures ($T_n$, $T_i$, and $T_e$) are compiled by post-processing the MS-MHD model outputs in the frame of AMPS. Details for the model parameters of MS-MHD are addressed in *Dong et al.* [2017a], and only descriptions that are relevant to this study are provided below.

We assume that PCb may have similar atmospheric compositions to Venus as PCb may have experienced a rapid loss of significant water and $H_2$ over time [*Barnes et al.,* 2016; *Ribas et al.,* 2016]. As in Venus, the thermosphere of PCb is assumed to be composed mainly of O and $CO_2$. Following the same procedure in *Dong et al.* [2017a], we use a scaled thermospheric profile of O and $CO_2$ based on PCb's mass and size [*Anglada-Escude et al.,* 2016] and the assumption of 1-bar surface pressure. Table 1 summarizes the planetary parameters and scaled neutral atmospheric profiles for O and $CO_2$. The MS-MHD model used one-third of the Earth dipole moment value with the same dipole polarity to describe the magnetized condition, consistent with certain dynamo scaling laws [*Dong et al.,* 2017a].

**Table 1.** The stellar wind parameters for Case 1 (C1) and Case 2 (C2). These parameters were used for the atmospheric ion loss study by *Dong et al.* [2017a], which provides the simulated

ionosphere to AMPS for this study. C1 and C2 describe the high and low stellar wind dynamic pressure conditions over one orbital period of PCb, respectively.

| Stellar wind parameters | $n_{SW}$ $(cm^{-3})$ | $T_{SW}$ $(K)$ | $v_{SW}$ $(km/s)$ | $B_{IMF}$ $(nT)$ |
|---|---|---|---|---|
| Case 1 (C1) | 21400 | $8.42 \times 10^5$ | (-833, 150, 0) | (0, 0, -227) |
| Case 2 (C2) | 2460 | $9.53 \times 10^5$ | (-1080, 150, 0) | (0, 0, -997) |
| *Neutral atmosphere* | | | | |
| $[CO_2] = 1.1 \times 10^{13} \times e^{-(z-z_0)/4.7}\ cm^{-3}$ | | | | |
| $[O] = 2.2 \times 10^9 \times e^{-(z-z_0)/14.5}\ cm^{-3}$ | | | | |

In this work, both unmagnetized and magnetized cases are studied. The first row of Figure 2 depicts the magnetic field configuration in the meridional plane for the magnetized (C1-M) and unmagnetized (C1-UnM) cases under the high stellar wind dynamic pressure conditions, respectively. The second row shows the corresponding cases under the low stellar wind dynamic pressure conditions (C2-M and C2-unM). One of striking difference between the first and second columns is that the polar regions become extended in the magnetized case, where planetary ions such as $O^+$ and $O_2^+$ can be transported to higher altitudes and may escape to space [*Dong et al., 2017a*].

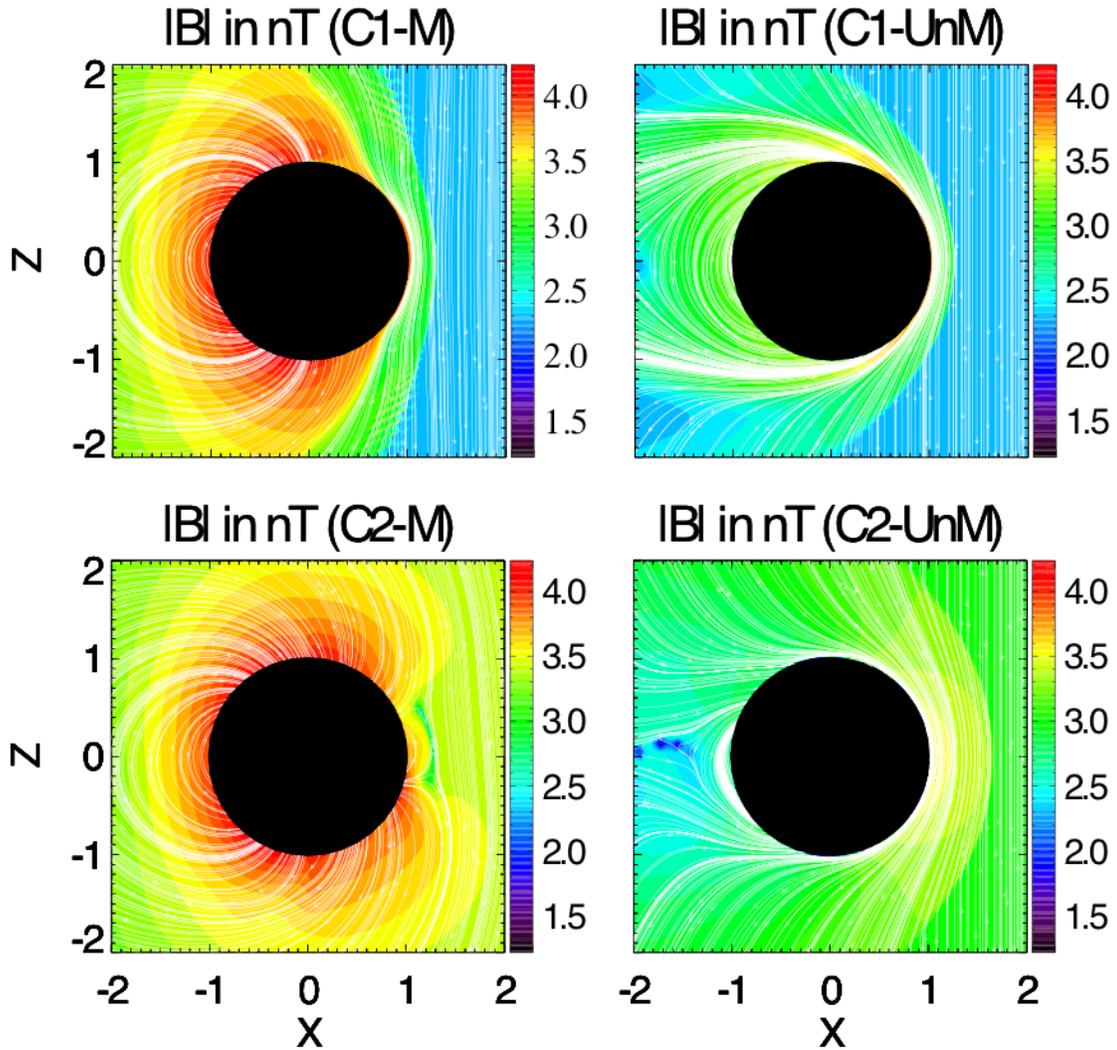

**Figure 2.** The logarithmic scale contour plots of the magnetic field strength (in nT) with magnetic field lines (in white) in the meridional plane for the magnetized case 1 (C1-M), unmagnetized case 1 (C1-UnM), magnetized case 2 (C2-M), and unmagnetized case 2 (C2-UnM).

In the model domain of AMPS, the production and loss of each model particle are described self-consistently. The model traces individual model particles once they are injected into the computational domain. As the model particles move under the influence of the planet's gravitational field, they undergo collisions with the ambient background particles, which modifies their position and momentum. After each collision, the model particle can either be thermalized and removed from the domain or completely escape to the space based on the local energy criterion of the model.

## 3. Formation of Hot Atomic Corona

### 3.1. Production of Hot Oxygen Atoms

At the present-day condition, it is widely accepted that the major source of hot O atoms is the dissociative recombination (DR) of $O_2^+$ with the ambient electron in the ionosphere of Venus and Mars (Eqn. 1) [*e.g., McElroy*, 1972; *Nagy and Cravens*, 1988; *Ip,* 1990; *Kim et al.,* 1998; *Krestyanikova and Shematovich*, 2005; *Chaufray et al.,* 2007; *Cipriani et al.,* 2007; *Fox and Hać*, 2009, 2014; *Valeille et al.*, 2009a, 2009b; *Yagi et al.*, 2012; *Lee et al.*, 2015a, 2015b; *Cravens et al.*, 2017; *Rahmati et al.*, 2017], where the thermospheres and ionospheres are composed mainly of $CO_2$ and $O_2^+$. Other minor sources of the O escape that are not considered here, such as sputtering by energetic pick-up ions and minor chemical reactions, can also contribute directly or indirectly to the production of escaping hot O atoms. This study considers DR of $O_2^+$ as the major source of hot O.

$$O_2^+ + e \rightarrow O(^3P) + O(^3P) + 6.99\ eV, \tag{1a}$$

$$\rightarrow O(^1D) + O(^3P) + 5.02\ eV, \tag{1b}$$

$$\rightarrow O(^1S) + O(^3P) + 2.80\ eV, \tag{1c}$$

$$\rightarrow O(^1D) + O(^1D) + 3.06\ eV, \tag{1d}$$

$$\rightarrow O(^1D) + O(^1S) + 0.83\ eV. \tag{1e}$$

As shown in Eqn. 1, the hot O corona structure is directly affected by the underlying ionospheric and thermospheric conditions and plasma temperatures. Specifically, the production rate of hot O depends on 1) the densities of parent ion and electron and 2) the rate coefficient of the reaction, where the latter is a function of electron temperature. Each channel in the reaction has the exothermicities. The branching ratio of each channel determines the resulting energy of hot O atoms. The two nascent hot O atoms share the excessive energy equally per the conservation of momentum and energy. This study adopts the branching ratio measured by *Kella et al.* [1997] for ground state $O_2^+$ ions and cold electrons with 0 eV relative collision energy. We do not include the vibrational or rotational energies of the parent ion, $O_2^+$, as most $O_2^+$ in the altitude range important for escaping hot O is in their ground state [*Fox and Hac,* 2009].

The nascent hot O undergoes several collisions with the ambient thermospheric species and may lose its energy and become a part of the thermosphere (thermalization process) before reaching the altitudes where collisions are infrequent. In addition to the DR of $O_2^+$, our model also considers a secondary production of hot O, which occurs via collisions with thermospheric O atoms (also called 'cold O'). The energy transferred from hot O to cold O during collisions is highly dependent on both the total and angular differential scattering cross-sections. When the energy of hot O transferred to cold O via collisions is enough to energize the cold O to become a new hot O atom, our model introduces a new particle to the domain as a collisionally-produced secondary hot O.

## 3.2. Ionosphere and Hot O Corona of Proxima b

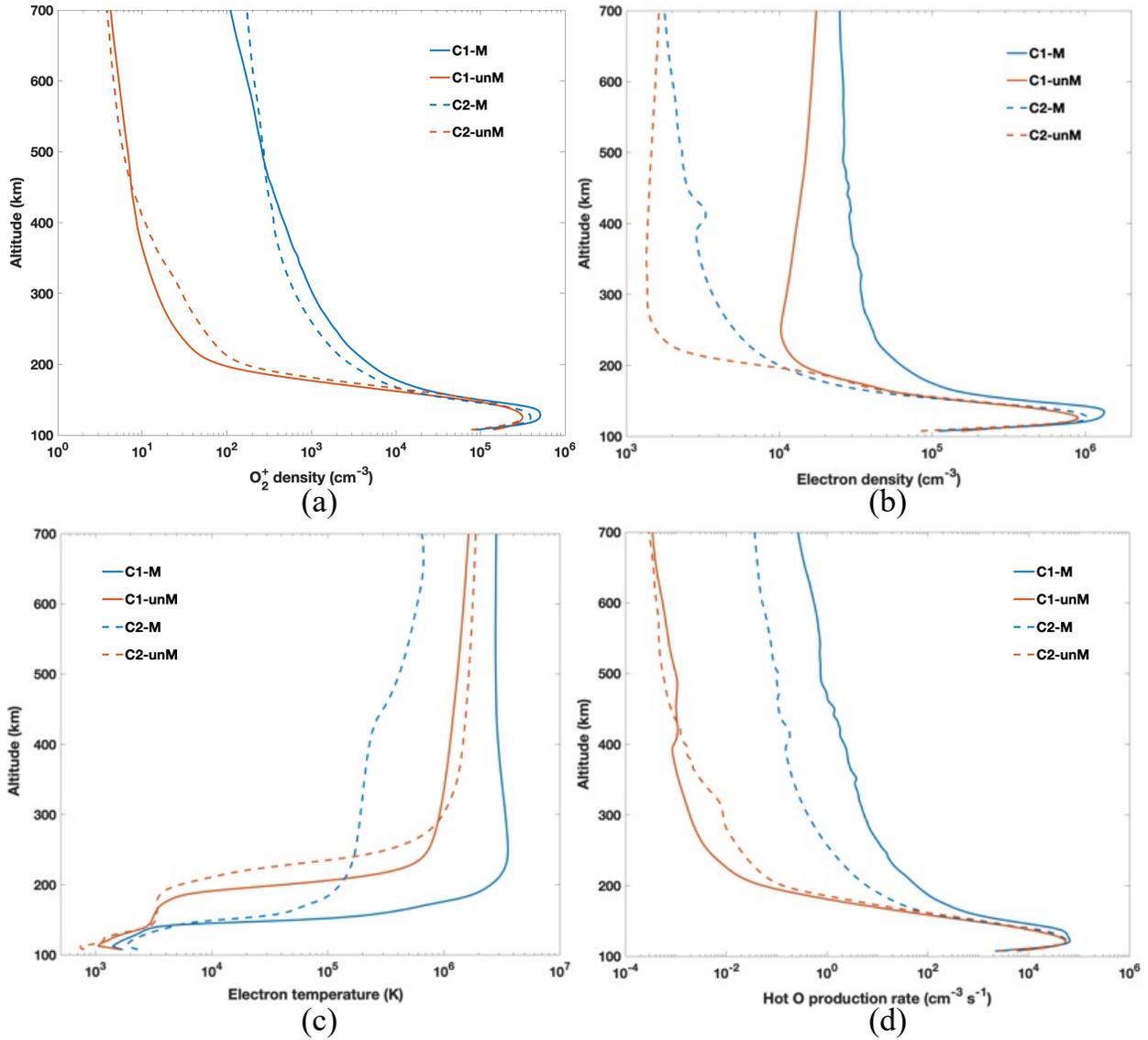

**Figure 3.** Altitude profiles of the globally-averaged a) $O_2^+$ density, b) electron density, c) electron temperature, and d) volume production rate of hot O for all four cases considered in this study. C1-M and C1-unM denote the magnetized and unmagnetized Case 1, respectively. Similar notations are also adopted for Case 2.

Figure 3 shows the globally-averaged altitude profiles of the $O_2^+$ density, electron density, electron temperature ($T_e$), and hot O production rate for the four cases. $O_2^+$ and electron both peak at an altitude deep in the ionosphere (~120 km) and decrease exponentially. The ionospheric structure below ~200 km is not affected significantly by either the variation in the stellar wind dynamic pressure or the presence of the intrinsic magnetic field. At higher altitudes, the ionospheric density become persistently more significant for all altitudes when the planet is magnetized, indicating that 1) the proton precipitation (along the open magnetic field lines) can charge exchange with the thermospheric species (at relatively low altitudes) near the polar regions and increase the ionospheric density, and 2) the ionosphere becomes extended with the presence of a global dipole magnetic field.

Figure 3b depicts that the electron density at higher altitudes (above ~200 km) for the high stellar wind dynamic pressure cases (C1-M and C1-unM) is much larger than that of the low stellar wind dynamic pressure cases (C2-M and C2-unM). This electron density difference at higher altitudes is associated with the different stellar wind (electron) density between C1 and C2 (see Table 1) as the electron density converges to that of the stellar wind with increasing altitude. The electron density difference at lower altitudes is consistent with the $O_2^+$ ion density since $O_2^+$ ion dominates the ionospheric peak in a $CO_2$- dominated atmosphere [*Dong et al*. 2017a]. It is noteworthy that the ionospheric density at higher altitudes is dominated by $O^+$ ion density owing to the larger scale height of O, leading to the $O^+$ density peaks at a higher altitude (> ~180 km) [*Dong et al.,* 2017a; Figure 2].

The electron temperature $T_e$ (in Figure 3c) changes dramatically in the ionosphere. For unmagnetized cases (C1-unM and C2-unM), $T_e$ gradually converges to the solar wind temperature (in Table 1) above 300 km. On the other hand, it requires higher altitudes (> 700 km) for $T_e$ to converge to the solar wind temperature under the magnetized cases due to the magnetic shielding effect. Compared with the low stellar wind dynamic pressure case (C2-M), $T_e$ of C1-M (see Figure 3c) is generally higher due to the relatively strong compression caused by the higher stellar wind dynamic pressure. However, $T_e$ of C2-M will be higher than that of C1-M once converging to the solar wind temperature (in Table 1) at higher altitudes.

The resulting volume production rates of hot O based on the densities of $O_2^+$ and electron and electron temperature are shown in Figure 3d. The hot O production is generally maximized deep in the ionosphere (~120 km), where the density of the parent ion, $O_2^+$, is also maximized and decreases with increasing altitude. The effects of the stellar wind dynamic pressure on the hot O production are more distinct when the planet is magnetized, mainly caused by a larger difference between the associated electron densities and temperatures, especially at high altitudes. In the magnetized scenario, the larger scale height of the ionosphere leads to an extended hot O production toward high altitudes compared to the unmagnetized scenarios. This effect depends on the topology of the planet's magnetosphere (see Figure 2), which requires a 3D analysis to better understand the structure of the entire process.

## 4. Results

### 4.1. Hot O Corona

Figure 4a displays the meridional and equatorial views of the simulated hot O corona of C2-unM only as those of other three cases show similar features. For all four conditions considered in this study, the overall structure of the hot O corona shows that the dayside envelope is radially inflated and decreases with increasing distance from the planet. The coronal density starts to decrease beyond the terminator and polar regions as the ionospheric density decreases in the planet's nightside due to the lack of an ionization source. The relatively shallower nightside corona is maintained primarily by two processes: 1) hot O transported from the dayside and 2) local production. If any hot O atoms traveling tailward near the planet's surface do not have enough energy to overcome the planet's gravitational force, they are attracted towards the planet and fall back to the upper atmosphere, constituting the nightside hot O corona [*Valeille et al.,* 2010a]. The factors that play an essential role in shaping the spatial structure of the hot O corona are 1) interaction between hot O and the ambient thermospheric particles via collisions and 2) generation of secondary hot O (*i.e.,* collisionally-produced hot O).

The effects of stellar wind dynamic pressure variation and magnetization of the planet on the hot O corona distribution are shown in Figures 4b – 4e. Figures 4b and 4c show the ratios of different stellar wind dynamic pressure conditions on the hot O corona density when the planet is assumed to be either unmagnetized or magnetized, respectively, and Figures 4d and 4e depict the ratios of magnetization on the hot O corona density under the high and low stellar wind dynamic pressure conditions, respectively. It must be noted that the ratios shown here are for providing the qualitative views of the effects in the near-planet region only. The hot O corona for the magnetized or high stellar wind dynamic pressure condition is more extended than their counterpart conditions, resulting in relatively large values in ratios at a greater distance from the planet where the coronal density decreases to a few cm$^{-3}$. To qualitatively understand the effects of different conditions in the near-planet environment, we added a trivial artificial value to the denominator when computing the ratios, which is large enough to dramatically decrease the ratio at a large distance (*i.e.*, beyond ~1 planet radius in altitude) but negligible to the coronal density near the planet. An inspection of Figure 4 reveals that magnetization of the planet leads to a relatively larger hot O density in the nightside. When the planet is magnetized, the ionospheric species, including $O_2^+$ (the parent ion of hot O), can be transported from the polar regions to the nightside along the dipole magnetic field lines. Especially, this effect is more enhanced under higher stellar wind dynamic pressure, resulting in a greater difference between the unmagnetized and magnetized conditions as shown in Figure 4d.

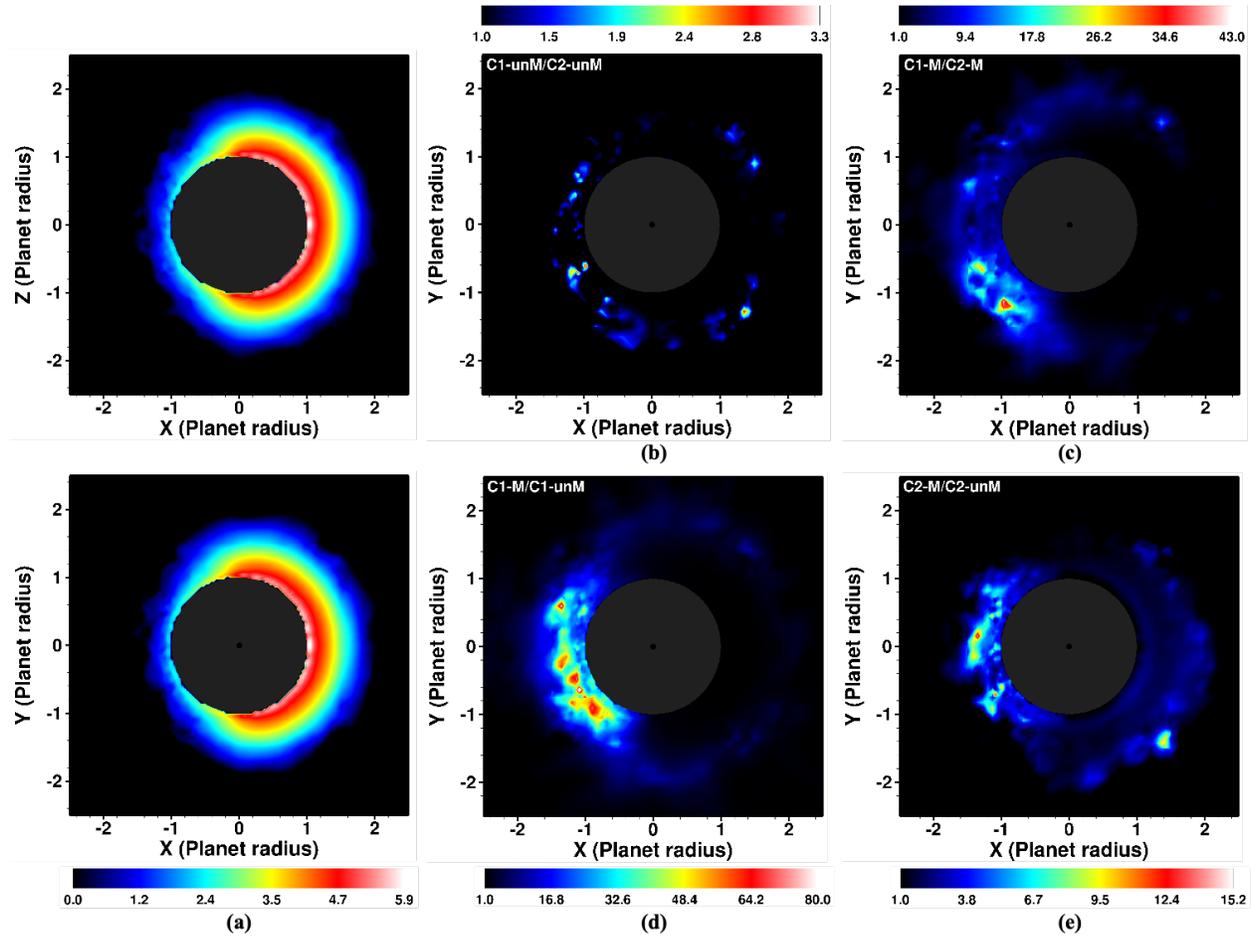

**Figure 4.** (a) Meridional (top) and equatorial (bottom) plane views of the hot O corona density for the C2-unM case. The star is located on the positive X-axis. The inner circle (dark gray) of each panel represents the planet. The color contour indicates the logarithmic scale of the hot O density (cm$^{-3}$). The purpose of showing only one case is to depict the overall features of the hot O corona envelope. The other three cases are not shown here as they share similar features. Equatorial plane

views of the ratios of the hot O density between (b) C1-unM and C2-unM, (c) C1-M and C2-M, (d) C1-M and C1-unM, and (e) C2-M and C2-unM are shown.

Figure 5 shows the globally averaged hot O density profiles from ~200 km to 3000 km altitude. The effect of the stellar wind dynamic pressure variation on the overall density structure of the hot O corona is not significant for the unmagnetized conditions but becomes essential for the magnetized state. Unlike the unmagnetized condition, the higher stellar wind dynamic pressure induces a considerable increase in the globally-averaged coronal density. The enhanced hot O density below ~600 km under higher stellar wind dynamic pressure conditions is mainly due to the enhanced ionospheric density by increased charge exchange with precipitation protons in the polar regions in the nightside. As altitude increases, the higher hot O density is more attributed to the higher electron density (see Figure 3b).

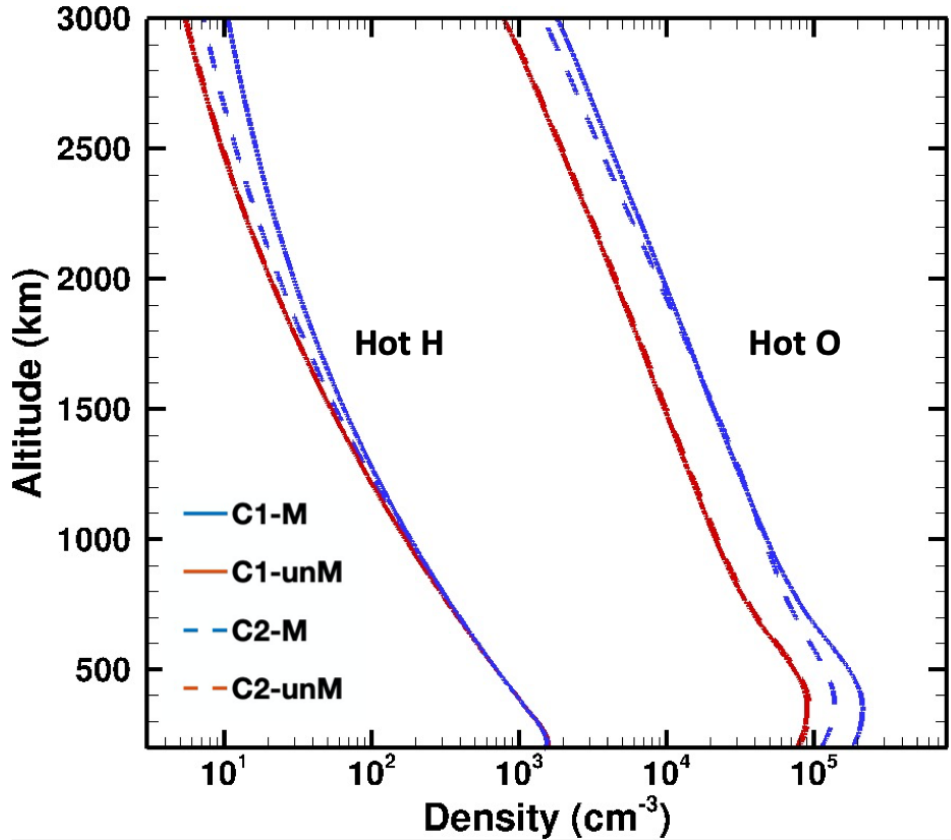

**Figure 5.** Globally-averaged altitude profiles for the simulated hot O and hot H densities. The profile sets on the left and right in the plot correspond to hot H and hot O, respectively.

**4.2. Potential Extended Hot H Corona**

The hot O corona simulated for this study is more extended and thicker than those on Mars and Venus, potentially a reservoir to collisionally-produced hot H. As shown in Eqn. 2, the resulting hot H is sensitive to the density and temperature of the underlying thermal (or cold) H and the hot O corona. Via elastic energy transfer collisions with the hot O atoms in the exosphere, the H atoms become energetic enough to escape the planetary atmosphere or form an extended H corona. Earlier studies have found that this energy transfer process is essential for the escape of H and D from Venus [*McElroy et al.,* 1982; *Rodriquez et al.,* 1984; *Shizgal and Arkos*, 1996].

$$H_{cold} + O_{hot} \rightarrow H_{hot} + O_{cold} \qquad (2)$$

To examine the potential hot H corona on the planet, we adapted the cold (thermal) H density at 200 km (approximate exobase altitude for our modeled PCb) for Venus estimated by *Fox and Sung* [2001]. The H density is scaled with the same assumptions for O and $CO_2$. In our AMPS model domain, we launched the cold H based on its density and temperature at 200 km into the background atmosphere composed of cold O and hot O, where the former is from our thermosphere profile and the latter is from our hot O corona simulation. We did not include $CO_2$ in the background atmosphere for this case study as the atmosphere above ~200 km is dominated by O.

The density profiles of the resulting hot H corona for all four cases are shown in Figure 5. Hot H is more extended to higher altitudes than thermal H, making the entire H exosphere more extended. The collisionally-produced hot H is mostly thermalized or, in other words, loses substantial energy due to collisions with thermal O deep in the thermosphere. As in the case of hot O, the magnetization of the planet is the dominant factor that affects the resulting structure of the H corona. In our model results, no difference between hot H at lower altitudes among the four cases is caused by the fact that collisions with thermal O dominate over hot O at lower altitudes and that the effects of thermal H collisions with hot O from different conditions become significant at high altitudes. For planets with their space environment similar to our modeled PCb, this process is expected to be efficient, indicating the potentially observable H exospheric envelope may be composed of a considerable amount of hot H.

### 4.3. Escape rates

**Table 2.** The calculated photochemical escape rate of hot O from our modeled PCb for all four cases considered in this study.

| O Escape rate ($s^{-1}$) | C1 | C2 |
|---|---|---|
| **Unmagnetized (unM)** | $2.7 \times 10^{22}$ | $1.8 \times 10^{21}$ |
| **Magnetized (M)** | $1.5 \times 10^{24}$ | $1.1 \times 10^{23}$ |

The escape rates for O computed by our model framework are summarized in Table 2. Among the four cases considered in this study, neutral O atoms escape to space the most for the C1-M case (*i.e.,* high stellar wind dynamic pressure and magnetized conditions). Regardless of stellar wind dynamic pressure, when the planet is magnetized, the escape rate is increased by about two orders of magnitude, primarily due to the enhanced hot O production from the extended ionosphere. The effect of the stellar wind dynamic pressure variation for both the unmagnetized and magnetized on the escape rate can change by about an order of magnitude. This increase is mainly attributed to the enhanced hot O production in the nightside under the high stellar wind conditions.

### 5. Discussion and Conclusion

We have simulated the exospheric structure of PCb to help us understand the effects of star-planet interaction on the upper atmosphere and atmospheric loss. The four cases considered in this study are designed to explore the effects of magnetization of the planet and variable stellar wind dynamic pressure on the escape of the neutral O atoms. The simulations are carried out by utilizing our model framework, which integrates the 3D exosphere, ionosphere, and magnetosphere models. As the atmosphere of PCb has not yet been observed, this study is entirely built upon the assumptions based on our Solar System's terrestrial planet, Venus, for the parameters relevant to PCb with the modeling technique established by our previous study on Mars. Therefore, this study is not intended to portray the actual exosphere and photochemical loss of PCb. Instead, one of the primary purposes of this study is to demonstrate the variabilities of the photochemical escape rate and exospheric structure of exoplanets that may have similar stellar and planetary conditions assumed here.

Given the nature of 3D models, this study provides a more realistic description of the escape process and a characterization of the uppermost layer of the planet's atmosphere resulting from its dual interactions with stellar winds and stellar radiation. While 1D models are computationally less expensive for exploring the parameter space, escape estimations for conditions where 3D dynamics are important, as shown in this study, can be aided by 3D models as a guideline to establish simplified assumptions.

The planetary energy balance is affected by cooling via the efficient escape of atmospheric constituents in the upper atmosphere. For our Venus-like Proxima b, the atmospheric loss predominantly occurs via ion loss, where neutral loss via the photochemical process adds a minor contribution. The hot atomic corona formed due to the nonthermal population contributes to the heating in the upper atmosphere, but the photochemical escape in our modeling condition does not lead to a significant effect on the planet's energy balance.

**1) Effects of Stellar wind dynamic pressure variation**

Our integrated modeling of the exosphere and nonthermal neutral O loss of PCb shows that if PCb possesses a Venus-like atmospheric composition, the planet will likely undergo considerable O loss in the form of neutral atoms via the photochemical process and form a more extended hot O corona than those on Venus. The resulting density structure of the hot O corona is not significantly affected by the variable stellar wind dynamic pressure, but the global hot O escape rate for the high stellar pressure condition (C1) is larger by one order of magnitude than that of the low stellar pressure condition (C2).

**2) Effects of planet's intrinsic magnetic field**

Both the structure and escape rate change dramatically when a global dipole field is imposed on the planet. When the planet is magnetized, the ionosphere becomes extended, leading to a more intensive hot O corona compared with those of the unmagnetized cases. The resulting global photochemical escape rate shows about two orders of magnitude increase compared to the unmagnetized condition. Unlike magnetization effects on the neutral escape, the atmospheric ion loss computed by *Dong et al.,* [2017a] shows an opposite behavior – high escape rate for the unmagnetized condition and reduced escape rate for the magnetized condition. In this regard, the atmospheric ion loss is closely related to the erosion by the stellar wind, whereas the atomic neutral loss is primarily dependent on the corresponding planet's ionospheric and atmospheric conditions. The neutral loss via photochemical mechanism depends on the resulting conditions of the ionosphere and thermosphere from the interaction between the upper atmosphere and stellar wind. This study finds that the photochemical escape is not the dominant atmospheric loss channel on PCb, based on our model setup, but becomes an essential contribution to the O loss when the planet is magnetized under the high stellar wind dynamic pressure condition.

**3) Implications to observations (Hot H corona)**

We have also demonstrated that the planet's hot O corona may play an essential role in a potentially detectable H exosphere. Via elastic collisions between thermal H and hot O, the planet's H corona may also have a two-component structure as the oxygen exosphere on Mars and Venus, where hot H at high altitudes leads to a more extended H exosphere. Although other sources of H may supply H to high altitudes, this study exemplifies that any future observations of the exosphere of PCb or similar M-dwarf exoplanets need to be interpreted with a consideration of contributions from non-thermal populations.

Given the highly variable and intense stellar wind environment of Proxima Centauri and the exosphere investigated in this study, other processes may also significantly affect the overall inventory of the PCb's atmosphere over the planet's history. For instance, the upper atmosphere of PCb may also be affected by the sputtering process (especially for the unmagnetized cases) and frequent bursts of large stellar flares.

**4) Implications to extreme space weather**

The stellar wind and upper atmosphere in this study are assumed to be the quiet and nominal conditions. However, the stars like Proxima Centauri are flaring stars [*Davenport et al.,* 2016]. Their stellar flare activity may be one of the most significant challenges to the habitability of their planets as flares may strip their upper planetary atmospheres and damage the protective atmospheric layers such as the ozone layer at Earth. A single superflare can bring intense impacts to the planetary atmospheres, which can be recovered after a certain period [*e.g., Fang et al.*, 2019], but a highly frequent flare rate may permanently impact the atmosphere's composition, resulting in an inhospitable environment for organisms to survive. In addition to flaring activity, other disturbed stellar conditions such as stellar coronal mass ejection may also be sufficiently harmful to the planetary habitability and ultimately the evolution of the planetary atmosphere [*e.g., Jakosky et al.*, 2015; *Dong et al.*, 2015b, 2017b; *Luhmann et al.,* 2017]. We have learned from our Solar System how differently these events impact the planets with or without the intrinsic magnetosphere and different atmospheric composition. Applications of our Solar System examples to similar exoplanets will provide a baseline for improving our understanding of the terrestrial exoplanets, especially those in their close-in habitable zone, and assist the investigations of more exotic exoplanets.

**Acknowledgement**

We thank an anonymous reviewer for helpful comments on this manuscript. YL and CD were supported by NASA grant 80NSSC18K0288. CD was also supported by NASA grant 80NSSC21K0608 and Princeton Plasma Physics Laboratory. Resources for this work were provided by the NASA High-End Computing (HEC) Program through the NASA Advanced Supercomputing (NAS) Division at Ames Research Center. The Space Weather Modeling Framework that contains the AMPS and BATS-R-US codes used in this study is publicly available from http://csem.engin.umich.edu/tools/swmf. For distribution of model results in this study, please contact the corresponding author.

# References

Airapetian, V. et al. (2020). Impact of space weather on climate and habitability of terrestrial-type exoplanets. *International Journal of Astrobiology*, 19, 136-194.

Anglada-Escudé, G., Amado, P., Barnes, J. et al. (2016). A terrestrial planet candidate in a temperate orbit around Proxima Centauri. *Nature,* 536, 437–440, doi:10.1038/nature19106.

Barnes, R., et al. (2016). The habitability of Proxima Centauri b I: Evolutionary scenarios. arXiv:1608.06919.

Berger, M. J., and Colella, P. (1989). Local adaptive mesh refinement for shock hydrodynamics. *J. Comput. Phys.*, 82, 64-84.

Bockelée-Morvan, D., Debout, V., Erard, S., Leyrat, C., Capaccioni, F., and Filacchione, G. et al. (2015), First observations of H2O and CO2 vapor in comet 67P/Churyumov-Gerasimenko made by VIRTIS onboard Rosetta, *Astron. Astrophy.,* 583, A6. doi:10.1051/0004-6361/201526303.

Bonfils, X., et al. (2013). The HARPS search for southern extra-solar planets. XXXI. The M-dwarf sample. *Astron. Astrophys.,* 549, A109, doi:10.1051/0004-6361/201014704.

Carr, M.H. (1986). Mars: A water-rich planet. *Icarus* 68, 187–216.

Carr, M.H. (2007). The Surface of Mars. *Cambridge University Press*, London.

Catling, D. C., and Zahnle, K, J. (2009). The planetary air leak. *Sci. Am.,* 300(5):36-43, doi: 10.1038/scientificamerican0509-36.

Chabrier, G., and Baraffe, I. (2000). Theory of low-mass stars and substellar objects, *Ann. Rev. Astron. Astrophys.,* 38, 337.

Chassefière, E., et al. (2013). The fate of early Mars' lost water: The role of serpentinization. J. *Geophys. Res. Planets,* 118, 1123-1134, doi:10.1002/jgre.20089.

Chaufray, J. Y., R. Modolo, F. Leblanc, G. Chanteur, R. E. Johnson, and J. G. Luhmann (2007). Mars solar wind interaction: Formation of the Martian corona and atmospheric loss to space. *J. Geophys. Res.,* 112, E09009, doi:10.1029/2007JE002915.

Cipriani, F., F. Leblanc, and J. J. Berthelier (2007). Martian corona: Nonthermal sources of hot heavy species. *J. Geophys. Res.,* 112, doi:10.1029/2006JE002818.E07001.

Cohen, O., et al. (2015). The interaction of Venus-like, M-dwarf planets with the stellar wind of their host star. *Astrophys. J.,* 806, 41.

Combi, M., Tenishev, V., Rubin, M., Fougere, N., and Gombosi, T. (2012), Narrow dust jets in a diffuse gas coma : A natural product of small active regions on comets. *Astrophys. J.,* 749(1), 29. doi:10.1088/0004-637x/749/1/29.

Cravens, T. E., A. Rahmati, Jane L. Fox, R. Lillis, S. Bougher, J. Luhmann, S. Sakai, J. Deighan, Yuni Lee, M. Combi, and B. Jakosky (2017). Hot oxygen escape from Mars: Simple scaling with solar EUV irradiance. *J. Geophys. Res.*, 122, 1102, doi:10.1002/2016JA023461.

Davenport, J. R., et al. (2016). Most observations of our nearest neighbor: Flares on Proxima Centauri. *Astrophys. J.,* 829, L31.

Dong, C., et al. (2015a). Solar wind interaction with the Martian upper atmosphere: Crustal field orientation, solar cycle and seasonal variations. *J. Geophys. Res. Space Physics.,* 120, 7857--7872, doi:10.1002/2015JA020990.

Dong, C., et al. (2015b). Multifluid MHD study of the solar wind interaction with Mars' upper atmosphere during the 2015 March 8th ICME event, *Geophys. Res. Lett.*, 42, 9103-9112, doi:10.1002/2015GL065944.

Dong, C., et al. (2017a). Is Proxima Centauri b habitable? A study of atmospheric loss. *Astrophys. J. Lett.*, 837, L26, doi:10.3827/2041-8213/aa6438.

Dong, C., et al. (2017b). The dehydration of water worlds via atmospheric losses, *Astrophys. J. Lett.*, 847, L4.

Dong, C., et al. (2018a). Atmospheric escape from the TRAPPIST-1 planets and implications for habitability. *Proc. Natl. Acad. Sci.*, 115, 260-265, doi:10.1073/pnas.1708010115.

Dong, C., et al. (2018b), Modeling Martian Atmospheric Losses over Time: Implications for Exoplanetary Climate Evolution and Habitability, *Astrophys. J. Lett.,* 859, L14.

Dong, C., et al. (2018c), Solar wind interaction with the Martian upper atmosphere: Roles of the cold thermosphere and hot oxygen corona, *J. Geophys. Res. Space Physics* 123, 6639-6654, doi:10.1029/2018ja025543.

Dong, C., et al. (2019). Role of Planetary Obliquity in Regulating Atmospheric Escape: G-dwarf versus M-dwarf Earth-like Exoplanets. *Astrophys. J. Lett.*, 882, L16.

Dong, C., et al. (2020). Atmospheric Escape From TOI-700 d: Venus versus Earth Analogs. *Astrophys. J. Lett.*, 896, L24.

Dressing. C. D., and Charbonneau, D., (2015). The occurrence of potentially habitable planets orbiting M dwarfs estimated from the full Kepler dataset and an empirical measurement of the detection sensitivity. *Astrophys. J.,* 807, 45, doi:10.1088/0004-637X/807/1/45.

Ekenbäck, A., et al. (2010). Energetic neutral atoms around HD 209458b: Estimations of magnetospheric properties. *Astrophys. J.,* 709, 670-679.

Erkaev, N. V., et al. (2007). Aspects of solar wind interaction with Mars: comparison of fluid and hybrid simulations. *Ann. Geophys.,* 25, 145-159.

Erkaev, N. V., et al. (2013). XUV-Exposed, non-hydrostatic hydrogen-rich upper atmospheres of terrestrial planets. Part I: Atmospheric expansion and thermal escape. *Astrobiology.* 13, 11, doi:10.1089/ast.2012.0957.

Fang, X., et al. (2019). Mars upper atmospheric responses to the 10 September 2017 solar flare: A global, time-dependent simulation, *Geophys. Res. Lett.*, 46, 9334-9343, doi:10.1029/2019GL084515.

Fougere, N., Altwegg, K., Berthelier, J., Bieler, A., Bockelée-Morvan, D., & Calmonte, U. et al. (2016a), Three-dimensional direct simulation Monte-Carlo modeling of the coma of comet 67P/Churyumov-Gerasimenko observed by the VIRTIS and ROSINA instruments on board Rosetta. *Astron. Astrophys.,* 588, A134. doi:10.1051/0004-6361/201527889.

Fougere, N., Altwegg, K., Berthelier, J., Bieler, A., Bockelée-Morvan, D., and Calmonte, U. et al. (2016b), Direct Simulation Monte Carlo modelling of the major species in the coma of comet 67P/Churyumov-Gerasimenko. *Mon. Not. Roy. Astr. Soc.,* 462(Suppl 1), S156- S169. doi:10.1093/mnras/stw2388.

Wachmann 3 B. *Icarus,* 221(1), 174-185. doi:10.1016/j.icarus.2012.07.019.

Fox, J. L., and Hać, A. B. (2009). Photochemical escape of oxygen from Mars: A comparison of the exobase approximation to a Monte Carlo method. *Icarus*, 204, 527-544, doi:10.1016/j.icarus.2009.07.005.

Fox, J. L., and A. B. Hać (2014). The escape of O from Mars: Sensitivity to the elastic cross sections. *Icarus*, 228, 375-385, doi:10.1016/j.icarus.2013.10.014.

Fox, J. L., and Sung, K. Y. (2001). Solar activity variations of the Venus thermosphere/ionosphere. *J. Geophys. Res.,* 106, 21, 305–21,335.

Garraffo, C., et al. (2016). The space weather of Proxima Centauri b. *Astrophys. J. Lett.*, 833, L4.

Ip, W.-H. (1990). The fast atomic oxygen corona extent of Mars. *Geophys. Res. Lett.,* 17, 2289–2292.

Jakosky, B. M., et al. (2015). MAVEN observations of the response of Mars to an interplanetary coronal mass ejection, *Science,* 350, aad0210, doi:10.1126/science.aad0210.

Jakosky, B. M., et al. (2018). Loss of the Martian atmosphere to space: Present-day loss rates determined from MAVEN observations and integrated loss through time. *Icarus,* 315, 146-157.

Jakosky, B.M., and Phillips, R.J. (2001). Mars' volatile and climate history. *Nature,* 412, 237–244.

Kella, D., et al. (1997). The source of green light emission determined from a heavy-ion storage ring experiment. *Science,* 276, 1530–1533.

Kim, J., Nagy, A.F., Fox, J.L., Cravens, T.E. (1998). Solar cycle variability of hot oxygen atoms at Mars. *J. Geophys. Res.,* 103, 29339–29342.

Kislyakova, K. G., et al. (2013). XUV-exposed, non-hydrostatic hydrogen-rich upper atmospheres of terrestrial planets. Part II: Hydrogen coronae and ion escape. *Astrobiology,* 13, 11.

Kislyakova, K. G., et al. (2014). Stellar wind interaction and pick-up ion escape of the Kepler-11 "super-Earths". *Astron. Astrophys.,* 562, A116.

Kite, E. S., and Barnett, M. N. (2020). Exoplanet secondary atmosphere loss and revival. *PNAS,* 117 (31), 18264-18271, doi: 10.1073/pnas.2006177117.

Krestyanikova, M.A., Shematovich, V.I. (2005). Stochastic models of hot planetary and satellite coronas: A photochemical source of hot oxygen in the upper atmosphere of Mars. *Solar Syst. Res.,* 39, 22–32.

Lammer, H., et al. (2003). Estimation of the past and present Martian water-ice reservoirs by isotopic constraints on exchange between the atmosphere and the surface. *Int. J. Astrobiol.,* 2, 195–202.

Lammer, H., et al. (2009). What makes a planet habitable?. *Astron. Astrophys. Rev.,* 17:181-249, doi:10.1007/s00159-009-0019-z.

Lee, Y., Combi, M., Tenishev, V., Bougher, S., and Lillis, R. (2015a). Hot oxygen corona at Mars and the photochemical escape of oxygen: Improved description of the thermosphere, ionosphere, and exosphere. *J. Geophys. Res. Planets*, 120(11), 1880-1892. doi:10.1002/2015je004890.

Lee, Y., et al. (2015b), A comparison of 3-D model predictions of Mars' oxygen corona with early MAVEN IUVS observations, *Geophys. Res. Lett.*, 42, doi:10.1002/2015GL065291.

Lee, Y., M. R. Combi, V. Tenishev, and S. W. Bougher (2014a), Hot carbon corona in Mars' upper thermosphere and exosphere: 1. Mechanisms and structure of the hot corona for low solar activity at equinox, *J. Geophys. Res. Planets*, 119, 905–924, doi:10.1002/2013JE004552.

Lee, Y., M. R. Combi, V. Tenishev, and S. W. Bougher (2014b), Hot carbon corona in Mars' upper thermosphere and exosphere: 2. Solar cycle and seasonal variability, *J. Geophys. Res. Planets*, 119, 2487–2509, doi:10.1002/2014JE004669.

Lee, Y. et al., (2018), Effects of a Solar Flare on the Martian Hot O Corona and Photochemical Escape, *Geophys. Res. Lett.,* 45, 6814-6822, doi:10.1029/2018GL077732.

Lee, Y., et al. (2020), Effects of Global and Regional Dust Storms on the Martian Hot O Corona and Photochemical Loss. *J. Geophys. Res. Space Physics*, 125, e2019JA027115, doi:10.1029/2019JA027115.
32

Laneuville, M., C. Dong, J. G. O'Rourke, A. C. Schneider (2020), Magnetic Fields on Rocky Planets, Chapter in the Book "Planetary Diversity", AAS | IOP Astronomy. https://doi.org/10.1088/2514-3433/abb4d9ch3

Lopez, E. D. (2017). Born dry in the photo-evaporation desert: Kepler's ultra-short-period planets formed water-poor. *Mon. Not. R. Astron. Soc.,* 472, 1, doi:10.1093/mnras/stx1558.

Lopez, E. D., et al. (2012). How thermal evolution and mass-loss sculpt population of super-Earths and sub-Neptunes: Application to the Kepler-11 system and beyond. *Astrophys. J.,* 761:59.

Luhmann, J. G., et al. (2017). Martian Magnetic Storms, *J. Geophys. Res. Space Physics,* 122, 6185-6209, doi:10.1002/2016JA023513.

Ma, Y., et al. (2013). A global multispecies single-fluid MHD study of the plasma interaction around Venus, *J. Geophys. Res. Space Physics,* 118, 321-330, doi: 10.1029/2012JA018265.

McElroy, M. B., et al. (1982). Escape of hydrogen from Venus. *Science,* 215, 4540, 1614-1615, doi:10.1126/science.215.4540.1614.

McElroy, M.B. (1972). Mars: An evolving atmosphere. *Science,* 175, 443–445.

Meadows, V. S., et al. (2018). The habitability of Proxima Centauri b: Environmental states and observational discriminants. *Astrobiology*, 133-189, doi:10.1089/ast.2016.1589.

Nagy, A., Cravens, T.E. (1988). Hot oxygen atoms in the upper atmospheres of Venus and Mars. *Geophys. Res. Lett.,* 15, 433–435.

Owen, J. E., and Jackson, A. P. (2012). Planetary evaporation by UV and X-ray radiation: basic hydrodynamics. *Mon. Not. R. Astron. Soc.,* 425, 2931-2947.

Rahmati, A., D. E. Larson, T. E. Cravens, R. J. Lillis, J. S. Halekas, J. P. McFadden, P. A. Dunn, D. L. Mitchell, E. M. B. Thiemann, F. G. Eparvier, G. A. DiBraccio, J. R. Espley, C.

Mazelle, and B. M. Jakosky (2017). MAVEN measured oxygen and hydrogen pick-up ions: Probing the Martian exosphere and neutral escape. *J. Geophys. Res.*, 122, 3689, doi:10.1002/2016JA023371.

Ribas, I., et al. (2016). The habitability of Proxima Centauri b. *Astron. Astrophys.,* 596, A111.

Rodriquez, J. M., et al. (1984). Hydrogen on Venus: Exospheric distribution and escape. *Planet. Space Sci.,* 32, 10, 1235-1255.

Shizgal, B. D., and Arkos, G. G. (1996). Nonthermal escape of the atmospheres of Venus, Earth, and Mars. *Rev. Geophys.,* 34, 4, 483-505.

Tenishev, V., et al. (2021). Application of the Monte Carlo method in modeling dusty gas, dust in plasma, and energetic Ions in planetary, magnetospheric, and heliospheric environments. *J. Geophys. Res. Space Physics, 126,* e2020JA02824,. doi:10.1029/2020JA028242.

Tenishev, V., M. R. Combi, and B. Davidsson (2008), A Global kinetic model for cometary comae. The evolution of the coma of the Rosetta target comet Churyumov–Gerasimenko throughout the mission, *Astrophys. J.*, 685, 659–677.

Tenishev, V., M. R. Combi, and M. Rubin (2011), Numerical simulation of dust in a cometary coma: Application to comet 67P/Churyumov-Gerasimenko, *Astrophys. J.*, 732, 104, doi:10.1088/0004-637X/732/2/104.

Tenishev, V., M. Rubin, O. J. Tucker, M. R. Combi, and M. Sarantos (2013), Kinetic modeling of sodium in the lunar exosphere, Icarus, 226, 1538–1549, doi:10.1016/j.icarus.2013.08.021.

Tenishev, V., Öztürk, D., Combi, M., Rubin, M., Waite, J., and Perry, M. (2014). Effect of the Tiger Stripes on the water vapor distribution in Enceladus' exosphere. *J. Geophys. Res.,* 119(12), 2658-2667. doi:10.1002/2014je004700.

Tian, F. (2015). Atmospheric Escape from Solar System Terrestrial Planets and Exoplanets. Annual Review of Earth and Planetary Sciences, 43, 459-476.

Tóth, G. et al. (2012). Adaptive numerical algorithms in space weather modeling. *Journal of Computational Physics* 231 (3), 870-903.

Tuomi, M., et al. (2014). Bayesian search for low-mass planets around nearby M dwarfs – estimates for occurrence rate based on global detectability statistics. *Mon. Not. R. Astron. Soc.,* 441, 1545-1569.

Tuomi, M., et al. (2019). Frequency of planets orbiting M dwarfs in the Solar neighbourhood. Submitted to AAS journals, arXiv:1906.04644v2 [astro-ph.EP].

Valeille, A., V. Tenishev, S. W. Bougher, M. R. Combi, and A. F. Nagy (2009a). Three-dimensional study of Mars upper thermosphere/ionosphere and hot oxygen corona: 1. General description and results at equinox for solar low conditions. *J. Geophys. Res.,* 114, E11005, doi:10.1029/2009JE003388.

Valeille, A., M. R. Combi, S. W. Bougher, V. Tenishev, and A. F. Nagy (2009b). Three-dimensional study of Mars upper thermosphere/ionosphere and hot oxygen corona: 2. Solar cycle, seasonal variations, and evolution over history. *J. Geophys. Res.,* 114, E11006, doi:10.1029/2009JE003389.

Valeille, A., M. R. Combi, V. Tenishev, S. W. Bougher, and A. F. Nagy (2010a). A study of suprathermal oxygen atoms in Mars upper thermosphere and exosphere over the range of limiting conditions. *Icarus*, 206, 18-27, doi:10.1016/j.icarus.2008.08.018.

Valeille, A., M. R. Combi, V. Tenishev, S. W. Bougher, and A. F. Nagy (2010b). Water loss and evolution of the upper atmosphere and exosphere over martian history. *Icarus*, 206, 28-39, doi:10.1016/j.icarus.2009.04.036.

Waite, Jr., et al. (2006). Cassini ion and neutral mass spectrometer: Enceladus plume composition and structure. *Science,* **311**, 1419.

Winters, J. G., et al. (2015). The Solar neighborhood. XXXV. Distances to 1404 M dwarf systems within 25 pc in the southern sky. *Astrophys. J.,* 149:5, doi:10.1088/0004-6265/149/1/5.

Yagi, M., F. Leblanc, J. Y. Chaufray, F. Gonzalez-Galindo, S. Hess, and R. Modolo (2012). Mars exospheric thermal and non-thermal components: Seasonal and local variations. *Icarus*, 221, 682-693.